# Learning from Literature: Integrating LLMs and Bayesian Hierarchical Modeling for Oncology Trial Design

Guannan Gong[1], PhD; Satrajit Roychoudhury[2], PhD; Allison Meisner[3], PhD, Lajos Pusztai[1], MD, PhD; Sarah B Goldberg[1], MD, Wei Wei[1], PhD*

[1]Yale Cancer Center, Yale School of Medicine, New Haven, CT, USA
[2]Pfizer Inc, New York, NY, USA
[3]Public Health Sciences Division, Fred Hutchinson Cancer Center

*Corresponding Author: Wei Wei, PhD

Yale Cancer Center, Yale School of Medicine

New Haven, CT 06520, USA

Email: wei.wei@yale.edu

Keywords: oncology trial design, Bayesian hierarchical modelling, evidence synthesis, large language models

## Abstract

**Purpose**

Designing modern oncology trials requires synthesizing evidence from prior studies for hypothesis generation and sample size calculation. Trial designs built on incomplete information or imprecise summaries can lead to mis-specified hypotheses and underpowered studies, resulting in missed opportunities and false positive or negative trial results. To address this challenge, we developed LEAD-ONC (Literature to Evidence for Analytics & Design in Oncology), an AI-assisted platform that ingests results of published trials, converts unstructured data to quantitative information, and performs Bayesian evidence synthesis to support well-informed design decisions for future trials. As LEAD-ONC remains under active development, the analyses presented here are intended as preliminary demonstrations of the framework's potential for trial design rather than definitive clinical conclusions. The current version of LEAD-ONC is available at the developer's website (http://34.55.33.152/step1#).

**Methods**

Given expert-curated publications that meet prespecified eligibility criteria, the proposed framework uses large language models (LLMs) to extract summaries of baseline characteristics from "Table 1" and reconstruct individual-patient data (IPD) from Kaplan–Meier (KM) curves. Trials are then clustered by similarity in baseline characteristics to define subpopulations more comparable for evidence synthesis. Within a cluster mostly resembling the target trial population, we fit a Bayesian hierarchical model to the reconstructed IPD and generate the predictive distribution of survival curves for the target trial.

**Results**

We demonstrate the utility of LEAD-ONC using results from five recently completed phase III trials of non-small-cell lung cancer (NSCLC) testing PD-L1/PD-1 inhibitors with or without CTLA-4 inhibitors (i.e., trials of dual or mono Immune Checkpoint Inhibitors (ICIs) versus chemotherapy alone). These trials were selected due to overlapping key eligibility criteria. Using baseline characteristics extracted from selected publications, LEAD-ONC identified three distinct patient populations primarily defined by histology including all-non-squamous (KEYNOTE-189), all-squamous (KEYNOTE-407), and mixed-histology (POSEIDON, CheckMate-227, CheckMate-9LA). For a prospective randomized trial in the mixed-histology population comparing mono versus dual ICIs, LEAD-ONC predicted a low probability of clinically meaningful benefit for dual therapy: median overall survival (OS) difference is 2.8 months (95% CI, −2.0 to 7.6) and the probability that dual-ICIs improves median OS by at least 3 months was approximately 0.45.

**Conclusion**

The combination of LLMs and Bayesian hierarchical modeling enables effective learning from literature and provides an evidence-driven approach for the planning of future trials.

**Introduction**

Designing clinical trials requires the systematic synthesis of a large and rapidly expanding body of prior evidence. At the trial design stage, investigators routinely review numerous publications and draw on results from historical trials to inform critical decisions, including anticipated treatment effect size, required follow-up to accrue sufficient events, and eligibility criteria defining the target population. In practice, effective learning from prior trials is frequently hindered by limited access to individual patient data (IPD),[1] the labor-intensive and error-prone nature of extracting quantitative information from the published literature, and subjectivity in selecting and combining evidence across studies. Together, these limitations lead to imprecise or overly optimistic assumptions about key design parameters, increasing the risk of underpowered studies and costly failures in oncology drug development when expected outcome targets are not achieved.

To address these challenges, we developed LEAD-ONC, an AI-assisted methodological framework that transforms investigator-curated oncology trial publications into quantitative evidence to inform the design of a future trial, hereafter referred to as the target trial. To begin with, investigators apply prespecified eligibility criteria (population, treatment strategy, trial conduct, etc) to identify relevant studies and upload their selected publications to LEAD-ONC. Given this curated input, LEAD-ONC (Figure 1) uses large language models (LLMs) and optical character recognition to extract quantitative information from unstructured sources, including Kaplan–Meier (KM) curves and baseline covariates summaries ("Table 1"), and converts them into analyzable datasets comprising reconstructed individual-level survival data and harmonized baseline covariates.

Because heterogeneity may persist even among trials meeting similar eligibility criteria, LEAD-ONC focuses on baseline characteristics that are commonly reported across the selected historical studies to quantify between-trial similarity and cluster trials into more comparable subpopulations. Within the cluster most closely resembling the target trial population, individual patient data (IPD) reconstructed from KM curves are fitted using a Bayesian Hierarchical Model (BHM) to generate the predictive distribution of survival curves for the target trial. This predictive distribution synthesizes survival outcomes and baseline characteristics across curated historical trials and represents the totality of evidence relevant to the target trial's survival endpoints.

Compared to existing tools that rely on manual cropping of figures and repeated point-and-click digitization of a KM curve,[2, 3] LEAD-ONC supports batch ingestion, multi-figure parsing, and automated association of curves with their corresponding risk tables. Compared to existing approaches in evidence synthesis that largely ignores the covariates information,[4–7] LEAD-ONC reduces the heterogeneity between trials via clustering based on published summaries of baseline covariates and allows the estimates from BHMs to be interpreted in the context of a specific target population. In contrast, evidence synthesis approaches based on a single BHM applied to all the historical trials may generate prior knowledge that is not specific to the trial of interest. By automating evidence extraction from prior publications in a scalable manner, LEAD-ONC improves prior evidence synthesis that is central to successful trial design.

In this manuscript, we describe the architecture of LEAD-ONC, including AI-assisted extraction of quantitative information from trial reports and a targeted BHM framework that learns survival experience from selected historical trials to inform the design of a future trial. We demonstrate how this approach could guide critical design decisions for a prospectively randomized trial in NSCLC.

**MOTIVATING CASE STUDIES**

Dual immune checkpoint blockade targeting CTLA-4 and PD-L1 or PD-1 has demonstrated superiority over platinum-based chemotherapy as first-line therapy for patients with metastatic non–small-cell lung cancer (NSCLC). However, whether dual immune checkpoint inhibition provides additional overall survival (OS) benefit beyond PD-L1 or PD-1 monotherapy remains an open and clinically important question.[8] To address this question, Federico et al. conducted a systematic review with selection criteria focusing on phase 3 RCTs enrolling patients with treatment-naive advanced NSCLC and comparison of a PD-L1 or PD-1 inhibitor, with or without platinum-based chemotherapy or a CTLA-4 inhibitor, versus platinum-based chemotherapy.[8] They identified several RCTs with overlapping eligibility criteria, including POSEIDON, CheckMate-227, CheckMate-9LA, Keynote-189 and Keynote-407.[9–15] We summarize the trial-level patient profiles for these trials in **Table 1**. The individual patient-level data on baseline characteristics were not available.

Suppose we are interested in evaluating the effect of dual versus mono immune checkpoint inhibition (ICI) on OS in a randomized controlled trial (RCT) enrolling treatment-naive patients with metastatic NSCLC. At the design stage, we seek to leverage all available information from relevant historical trials (baseline patient characteristics from "Table 1", KM curves on patient OS) to address the following critical design questions: (1) Are there any meaningful differences in patient characteristics among these selected historical trials, implying some are more relevant for trial design than others? (2) What are reasonable assumptions for OS for patients randomized to the mono- and dual-inhibitor arms in the future trial? (3) What survival benefit is expected for dual versus mono ICI therapy? (4) Given these assumptions, what is the probability of success for the planned trial? In this paper, we demonstrate how LEAD-ONC can answer these questions.

**Methods**

**AI-Assisted Extraction of Quantitative Trial Results from the Published Literature**

Tools such as IPDfromKM [2] and WebPlotDigitizer [3] have made IPD reconstruction increasingly feasible with reasonable fidelity. In practice, their use in trial design remains underexplored because of workflow and scalability barriers. First, most pipelines still require manual cropping of figures out of the source PDF before any digitization can begin. This step is error-prone (axes or legends may be clipped), time-consuming, and difficult to standardize across users or studies, limiting throughput when dozens of figures must be processed. Second, many tools rely on point-and-click digitization for each curve and manual transcription of risk tables; they typically operate on one curve from one figure at a time, with little or no support for batch ingestion, multi-figure parsing, or automated association of curves with their corresponding risk tables. As a result, scaling across many trials requires substantial coordinator time, introduces opportunities for transcription errors, and yields audit trails that are hard to reproduce.

In contrast, LEAD-ONC provides a more automated, AI-assisted framework for extracting quantitative information from published oncology trial reports. The platform is designed to process KM curves across multiple treatment arms and trials in a unified workflow, reducing manual intervention and improving scalability. Beyond KM curves, LEAD-ONC also extracts summaries of baseline characteristics from published "Table 1" data, enabling downstream assessment of trial comparability and more targeted evidence synthesis. Figure 1 summarizes the end-to-end LEAD-ONC workflow, from investigator-curated literature input to structured data generation for subsequent Bayesian analysis.

**User-facing workflow.** Investigators upload trial PDFs, select Kaplan–Meier (KM) figures and relevant tables (numbers-at-risk; baseline characteristics/Table 1), verify auto-recognized values, and complete a brief three-point KM calibration and trace. Each upload receives a persistent study identifier that preserves subfigure qualifiers (e.g., "Fig 1A", "CheckMate-227-B") to prevent cross-panel mixing.

**AI-assisted risk-table recognition (primary LLM path with OCR fallback).** Treatment-arm names, x-axis time points, and numbers-at-risk are extracted using a multimodal large language model (Google Gemini 2.0 Flash) with base64 image submission and a structured JSON schema (exact arm nomenclature; aligned arrays). Exponential-backoff retries ($\leq 5$ attempts; 3–48 s) mitigate transient errors. Returned JSON undergoes schema validation (types, equal-length arrays, monotonically non-increasing at-risk counts) and is presented for rapid human confirmation/correction. If validation or confidence thresholds are not met, an OCR fallback is triggered: (i) risk-table region detection below the x-axis (OpenCV) with keyword-guided fallback; (ii) ROI preprocessing (grayscale, adaptive threshold, de-skew, artifact cleanup); (iii) recognition with PaddleOCR (numeric-optimized; English) with Tesseract as secondary; (iv) table reconstruction by aligning columns to tick centers or projection peaks and grouping rows by line height; and (v) domain constraints (non-increasing at-risk counts; grid-snapped time when drift is small; equal-length arrays). When both paths succeed, a fusion/adjudication step computes cell-wise diffs; minor isolated discrepancies (≤±1) favor the LLM output, while larger conflicts or rule violations favor OCR. A side-by-side diff view supports user adjudication.

**Three-point calibration and human-supervised digitization.** Before digitization a human calibration step is required because automated axis detection (OCR/LLM + CV) remains brittle across heterogeneous KM layouts—axes may be skewed by scanning, unlabeled or partially labeled, use non-orthogonal grids, plot months vs weeks, or show overlapping curves and censored ticks; moreover, the visible x-axis maximum does not always equal the terminal month implied by the risk table. A three-point calibration lets the user quickly anchor scale and orientation without full axis parsing and works across these variants. Before tracing, users click three labeled ticks to define the pixel→data transform: origin (0, 0); rightmost labeled x-tick (max, 0), where max is the verified terminal month from the risk table; and top y-tick (0, 100). These non-collinear correspondences determine a 2D affine mapping

$$t = a u + b v + c, \; s = d u + e v + f$$

capturing translation, rotation, scale, and mild shear (robust to skew). Users then trace each KM curve on an HTML5 canvas; pixel traces are immediately transformed to calibrated (t,s) pairs and stored as JSON. Multi-arm panels are supported.

**Standardization and KM constraints.** All traced curves are resampled to 500 evenly spaced time points via piecewise-linear interpolation with canonical KM properties enforced: (1) survival = 100% at t=0 (inserted if absent), (2) strictly increasing time, and (3) monotonically non-increasing survival. Violations trigger in-app warnings for targeted correction prior to finalization.

**Arm mapping and de-duplication.** Curves are linked to risk-table arms using a hierarchical matcher: (1) user-confirmed mapping; (2) exact normalized string match; (3) fuzzy partial match with length-aware constraints (±5 characters to avoid false positives, e.g., matching "chemotherapy" to "nivolumab + chemotherapy"); and (4) fallback to curve-color label. Arm names are preserved verbatim from the input figure.

**Quality control and validation.** Automated checks verify: (i) one-to-one alignment between coordinate files (xy.csv, 500 rows/curve) and risk tables (risk_table.csv) by study–arm key; (ii) equal-length arrays across time points and at-risk counts; (iii) satisfaction of KM constraints (initialization and monotonicity); and (iv) uniqueness of study identifiers, including subfigure suffixes. Any failure blocks export and prompts targeted fixes.

**Individual patient data (IPD) reconstruction.** Taking the extracted XY coordinates (xy.csv) and the risk table (risk_table.csv) as input, individual patient data (IPD) associated with historical KM curves of a selected historical trial can be generated using the modified-iKM method[2].

**Assess Homogeneity of Selected Historical Trials**

Most publications of clinical trials include a Table 1 to summarize the distribution of key baseline characteristics: means and standard deviations for variables at a continuous scale, counts and percentages for categorical variables together with the sample sizes for each arm. Suppose we have selected $H$ historical trials that are relevant to the trial planned after applying a rigorous selection process, and we would like to evaluate the homogeneity of these $H$ historical trials and identify subpopulations by assessing the similarities in $p$ baseline covariates shared by these historical trials.

Let $x_{ij}$ and $x_{ik}$ denote the sample mean of the *i-th* baseline variable (e.g., gender, ECOG performance status) of two different trials indexed by $j$ and $k$. We use standardized difference as a measure of covariate dissimilarity[16]. For a continuous variable, the standardized difference between trials $j$ and $k$ is defined as

$$d_{ijk} = \frac{|x_{ij} - x_{ik}|}{\sqrt{\frac{S_{ij}^2 + S_{ik}^2}{2}}}$$

The corresponding sample variance $S_{ij}^2$ and $S_{ik}^2$ are commonly reported in "Table 1" of published trial results. For trials reporting median and range of continuous variables, we use the reported median to approximate the sample mean and approximate sample variance as the difference between maximum and minimum divided by four.

For binary variables, we consider

$$d_{ijk} = \frac{|x_{ij} - x_{ik}|}{\sqrt{\frac{x_{ij}(1-x_{ij}) + x_{ik}(1-x_{ik})}{2}}}$$

The standardized difference is commonly used as a measure of covariate balance in the matching and weighting literature, and it can be used as a balance metric across covariates measured in different units[16]. We define the dissimilarity in patient baseline profile between trials $j$ and $k$ by averaging the standardized differences associated with different covariates

$$D_{jk} = \frac{\sum_{1}^{p} d_{ijk}}{p}.$$

Alternatively, we can define the dissimilarity in patient baseline profile between two trials by taking the maximum of the standardized differences

$$D_{jk} = max(d_{1jk}, ..., d_{pjk}).$$

We can cluster trials using the K-medoid algorithm[17] or other method of choice based on $D_{jk}$ between pairs of historical trials. As a $d_{ijk}$ of *0.2, 0.5,* and *0.8* are generally used to represent small, medium, and large effect sizes[18], respectively, we consider a $D_{jk}$ (calculated either by taking the average or the maximum of standardized differences) of smaller than 0.2 as an indication of good comparability between trials, with larger values of $D_{jk}$ suggesting increased differences between trial populations.

**Evidence Synthesis Using Bayesian Hierarchical Modelling**

Bayesian Hierarchical Models (BHMs) such as the meta-analytic predictive prior (MAP) approach are commonly used for evidence synthesis when planning a trial.[6] Previous works in BHM largely focused on modeling different trial outcomes and often produce evidence that is hard to interpret as (1) different trials may have similar or different outcomes by coincidence and (2) it is not clear what patient population the synthesized evidence is derived from and how relevant the synthesized evidence is to a target population.

We choose to construct BHM in a cluster-specific fashion to ensure we can interpret the results from the BHM in the context of a specific target population. For trials in the cluster mostly resembling the target trial population, we construct a BHM assuming the survival functions of these trials are random samples from the same beta-Stacy process (BSP). The BSP model[19] consists of a precision function $c(t)$ and a parametric survival function $G(t)$, which is the mean of different survival functions associated with trials in the same cluster. Using reconstructed IPD from approximately 500 arms of phase 3 oncology trials, Deborah et al. showed a two parameter Weibull model can sufficiently describe survival curves from these arms[20]. Thus, we assume $G(t)$ is a Weibull model. We assume $c(t)$ equals to a constant $c$ such that information represented by a BHM is equivalent to $c$ patients. We assume independent noninformative priors for $c$ and parameters of $G(t)$. Details of the Bayesian Hierarchical Models can be found in the appendix.

Given the totality of evidence derived from prior trials, the posterior predictive distribution from the BHM represents the prior knowledge on the potential survival curves of the target trial of interest. **Figure 3** shows the predictive distribution of survival curves associated with mono and dural ICIs in a target trial enrolling patients with both squamous and non-squamous histology. This targeted evidence synthesis approach allows the user to select the cluster of greatest relevance and use the BHM results for that cluster to inform the planning of a target trial.

**Results**
**Baseline profiles and trial similarity**
**Table 1** summarizes the baseline profiles of patients enrolled in five phase III trials for treatment-naive advanced NSCLC. In CheckMate-227, patients were prospectively stratified by

PD-L1 expression into two trial parts. Patients with PD-L1 expression ≥1% were enrolled in Part 1a and randomized to receive nivolumab plus ipilimumab (dual therapy), nivolumab monotherapy, or chemotherapy (CT), whereas patients with PD-L1 expression <1% were enrolled in Part 1b and randomized to receive dual therapy, nivo+CT, or CT alone. Because the median OS was similar between the nivo arm in Part 1a and the nivo+CT arm in Part 1b (15.7 vs. 15.2 months), these two arms were pooled for downstream analysis. However, baseline characteristics for patients enrolled in Part 1b (PD-L1 <1%) were not reported by CheckMate-227. Consequently, we used baseline profiles of the dural therapy arm for clustering analysis.

We present the pairwise patient-profile dissimilarity of the five trials in **Figure 2**. The dissimilarity measures were calculated by taking the average (**Figure 2a:** left panel) or the maximum (**Figure 2b:** right panel) of the standardized differences associated with covariates commonly reported by these trials. K-medoids clustering yielded three groups primarily aligned with histology: non-squamous histology (KEYNOTE-189), squamous histology (KEYNOTE-407), and mixed histology consisting of CM227, CM9LA and POSEIDON. The cluster memberships for these trials were identical regardless of how the dissimilarity measures were calculated (average vs. maximum). Figure 2a shows that, on average, the standardized differences between covariates of two trials in the mixed-histology cluster is less than 0.20, whereas Figure 2b shows the standardized differences between covariates of two trials in the mixed-histology cluster is at most 0.22, indicating only modest differences across baseline covariates. Keynote-189 entirely consists of patients with non-squamous histology and Keynote-407 patients all have squamous histology. In addition, Keynote-189 includes more patients with ECOG=0 than other trials.

**Projecting survival for a new mixed-histology trial**
Each grey curve in **Figure 3** represents a possible survival curve that might be observed in a future randomized trial comparing mono– versus dual ICIs, assuming a patient baseline profile similar to that of trials in the mixed-histology cluster. Collectively, these curves characterize the uncertainty in the survival functions of the mono- and dual-inhibitor arms, as informed by Bayesian hierarchical modeling of reconstructed IPD from prior mono- and dual-inhibitor trials at the design stage. By averaging over these plausible survival curves, we obtain projected OS rates for each treatment arm, as well as projected differences in OS at prespecified time points, together with corresponding 95% credible intervals (CIs). **Table 2** summarizes the 95% CIs for projected OS differences (mono versus dual ICIs) across multiple time points. Because the credible intervals for OS differences at all evaluated time points include zero, the probability of demonstrating a clinically meaningful superiority of dual immune checkpoint inhibition in a future trial is limited under the assumed target population.

Based on the Bayesian hierarchical model fit to reconstructed IPD, the model-predicted median OS for the mono- and dual-ICI is 14.0 months (95% CI: 11.2-17.1) and 16.8 months (95% CI: 13.3-20.8), respectively. The projected difference in median OS is 2.8 months (95% CI: -2.0, 7.6). The projected probability of the dual therapy prolonging OS by at least 3 months compared to the mono therapy is approximately 0.45. Consistent with this finding, **Figure 4** shows substantial overlap in the 95% CIs of the projected medians for the two arms.

**Implications for a prospective RCT**
Taken together, these results suggest that, under the evidence synthesized from relevant historical trials of mono and dual immune checkpoint inhibition, the projected probability of success for a randomized trial comparing dual versus mono ICI in a population resembling the mixed-histology cluster is limited.

**Discussion**:
By integrating large language models with Bayesian hierarchical modeling, we developed LEAD-ONC, an end-to-end framework that transforms published oncology trials into quantitative, design-relevant evidence. Specifically, LEAD-ONC (i) quantifies comparability among historical trials using a dissimilarity metric derived from commonly reported baseline characteristics; (ii) converts published Kaplan–Meier survival curves into analyzable data through LLM-assisted extraction and individual patient data reconstruction; and (iii) generates posterior predictive distributions of survival functions for a prespecified target population using cluster-wise Bayesian hierarchical models. When applied to first-line NSCLC, clustering based on pairwise dissimilarity revealed a clinically interpretable structure, separating trials by histologic composition and clarifying which studies are most informative for specifying assumptions in a future trial.

Within the mixed-histology cluster (POSEIDON, CheckMate-227, and CheckMate-9LA), LEAD-ONC projected limited separation between survival curves for mono– versus dual–immune checkpoint inhibition. The estimated difference in median overall survival was modest, with substantial overlap in posterior credible intervals, indicating a limited probability of achieving a clinically meaningful benefit for dual blockade if evaluated in a future randomized trial enrolling a similar population. These findings illustrate how evidence extracted from the published literature, when combined with Bayesian modeling, can be used to formalize design-stage assumptions and support critical decision-making by explicitly quantifying uncertainty and the probability of trial success.

Oncology trials increasingly face escalating costs, prolonged timelines, and accrual challenges, particularly in settings where patient populations are fragmented by biomarker-defined subgroups. As a next step, we will evaluate how the proposed evidence synthesis framework can be extended to construct informative priors for use as synthetic controls in early-phase single-arm trials or to augment control arms in randomized studies, with the goal of improving efficiency and reducing required sample sizes while maintaining inferential rigor. Although this study focused on survival outcomes, the LEAD-ONC framework is general and can be readily extended to trials with binary or continuous endpoints.

At present, the metadata extraction and evidence synthesis modules of **LEAD-ONC** are implemented separately, and Bayesian hierarchical modeling for evidence synthesis is currently performed in a local computational environment. Ongoing development efforts are focused on expanding LEAD-ONC's functionality, including integration of advanced literature search capabilities, enhanced text and figure processing, and tighter coupling between extraction and Bayesian synthesis modules. These improvements are intended to support a more streamlined, scalable, and user-facing platform for evidence-driven trial design.

**Table 1**: Baseline characteristics of patients enrolled in five trials of treatment-naive advanced NSCLC. Summaries of baseline characteristics were extracted from these trials and merged based on covariates commonly available.

|  | Cluster 1 (Mixed Histology) | | | | Cluster 2 (Non-squamous) | Cluster 3 (Squamous) |
|---|---|---|---|---|---|---|
| Trial | CheckMate-9LA | CheckMate-227 | POSEIDON | | KEYNOTE-189 | KEYNOTE-407 |
| Arm | Nivo+Ipi (N=361) | Nivo+Ipi (N=583) | Tremelimumab+ Durvalumab+ CT (N=338) | Durvalumab +CT (N=338) | Pembrolizumab+ Chemo (n=410) | Pembrolizumab+ Chemo (N=278) |
| Age (Median, range) | 65(59-70) | 64(26-87) | 63(27-87) | 64.5(32-87) | 65(34-84) | 65(29-87) |
| Female(%) | 30 | 32.6 | 20.4 | 25.1 | 38 | 20.9 |
| ECOG (%) | | | | | | |
| 0 | 31 | 35 | 32.5 | 32.2 | 45.4 | 26.3 |
| 1 | 68 | 64.7 | 67.5 | 67.8 | 53.9 | 73.7 |
| Histology (%) | | | | | | |
| Squamous | 31 | 28 | 36.7 | 37.9 | 0 | 97.5 |
| Nonsquamous | 69 | 71.9 | 63.3 | 61.8 | 100 | 0 |
| Smoking status (%) | | | | | | |
| Never smoker | 13 | 13.6 | 17.5 | 24.9 | 11.7 | 7.9 |
| Former/Current smoker | 87 | 85.2 | 82.6 | 75.1 | 88.3 | 92.1 |
| PD-L1 >=1% (%) | 60 | 67.9 | 63 | 66.3 | 63.4 | 63.3 |

**Table 2:** Projected overall survival (OS) for the mono and dual ICIs and their projected differences at time points of interest in a new trial. These predictions are derived assuming the new trial has a baseline patient profile similar to trials in the mixed histology cluster (POSEIDON, CM227 and CM9LA).

| Time/month | OS (Mono-ICI) | | | OS (Dual-ICIs) | | | OS Differences | | |
| --- | --- | --- | --- | --- | --- | --- | --- | --- | --- |
| | Estimate | 2.5% CI | 97.5% CI | Estimate | 2.5% CI | 97.5% CI | Estimate | 2.5% CI | 97.5% CI |
| 12 | 0.54 | 0.48 | 0.59 | 0.58 | 0.52 | 0.63 | 0.04 | -0.04 | 0.12 |
| 24 | 0.35 | 0.30 | 0.41 | 0.41 | 0.36 | 0.46 | 0.06 | -0.02 | 0.13 |
| 36 | 0.24 | 0.20 | 0.29 | 0.30 | 0.25 | 0.36 | 0.06 | -0.01 | 0.13 |
| 48 | 0.17 | 0.13 | 0.22 | 0.23 | 0.19 | 0.28 | 0.06 | -0.01 | 0.12 |
| 60 | 0.13 | 0.09 | 0.17 | 0.18 | 0.14 | 0.23 | 0.05 | -0.01 | 0.12 |
| 72 | 0.09 | 0.06 | 0.13 | 0.14 | 0.10 | 0.19 | 0.05 | -0.01 | 0.10 |

**Figure 1**: LEAD-ONC workflow. LEAD-ONC transforms investigator-curated trial publications into quantitative, uncertainty-aware evidence to support the design of a prespecified target trial. (A) Investigators select relevant trial publications. (B) Large language models (LLMs) and optical character recognition are used to extract quantitative information from unstructured sources, including KM curves and baseline characteristic tables. (C) Extracted information is converted into reconstructed individual patient data and harmonized baseline covariates. (D) Trials are clustered according to similarity in baseline covariates. (E) Within the cluster most similar to the target trial population, Bayesian hierarchical models are used to synthesize evidence across trials. (F) The resulting predictive distributions for the survival curves of the target trial inform the choice of design parameters such as treatment effect size, and estimate the probability of trial success.

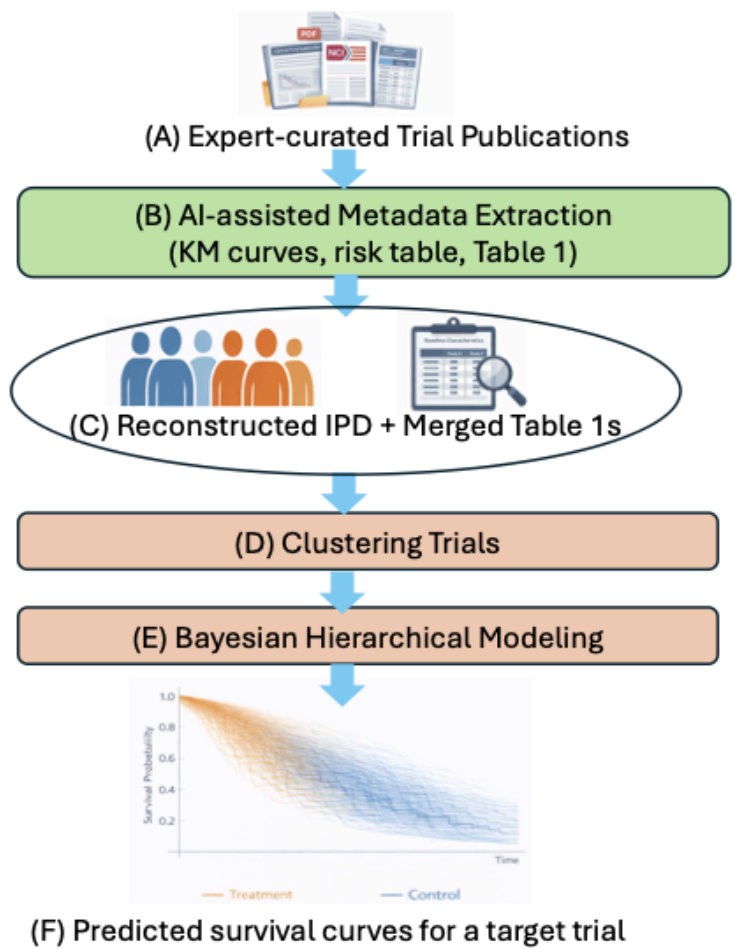

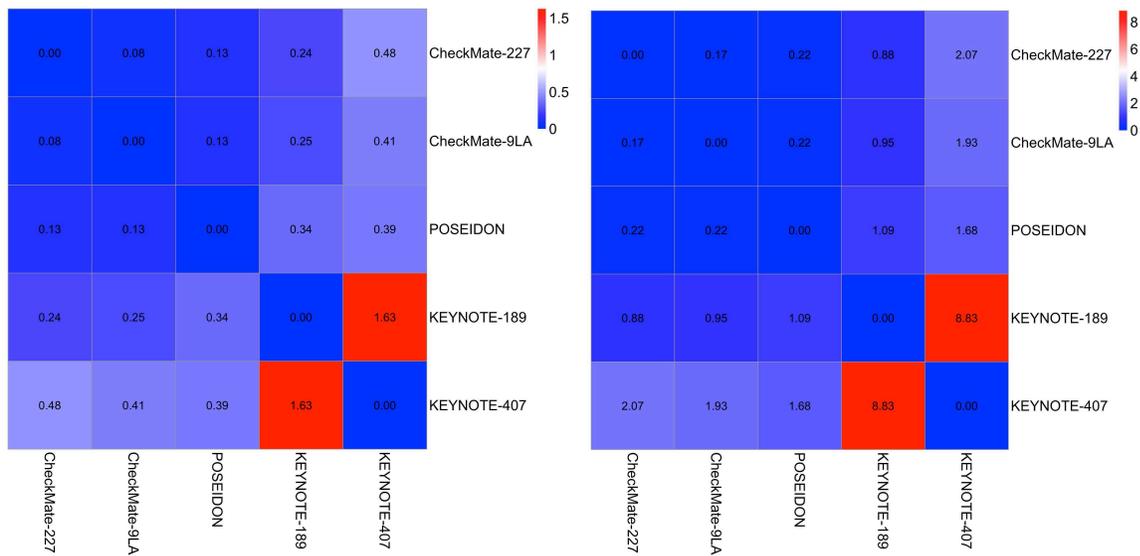

**Figure 2**: Dissimilarity of baseline patient profile between selected NSCLC trials, with values of 0.2, 0.5, and 0.8 representing small, medium, and large differences, respectively. Dissimilarity values were calculated by taking the average (Figure 2A: left panel) or the maximum (Figure 2B: right panel) of the standardized differences associated with all covariates available.

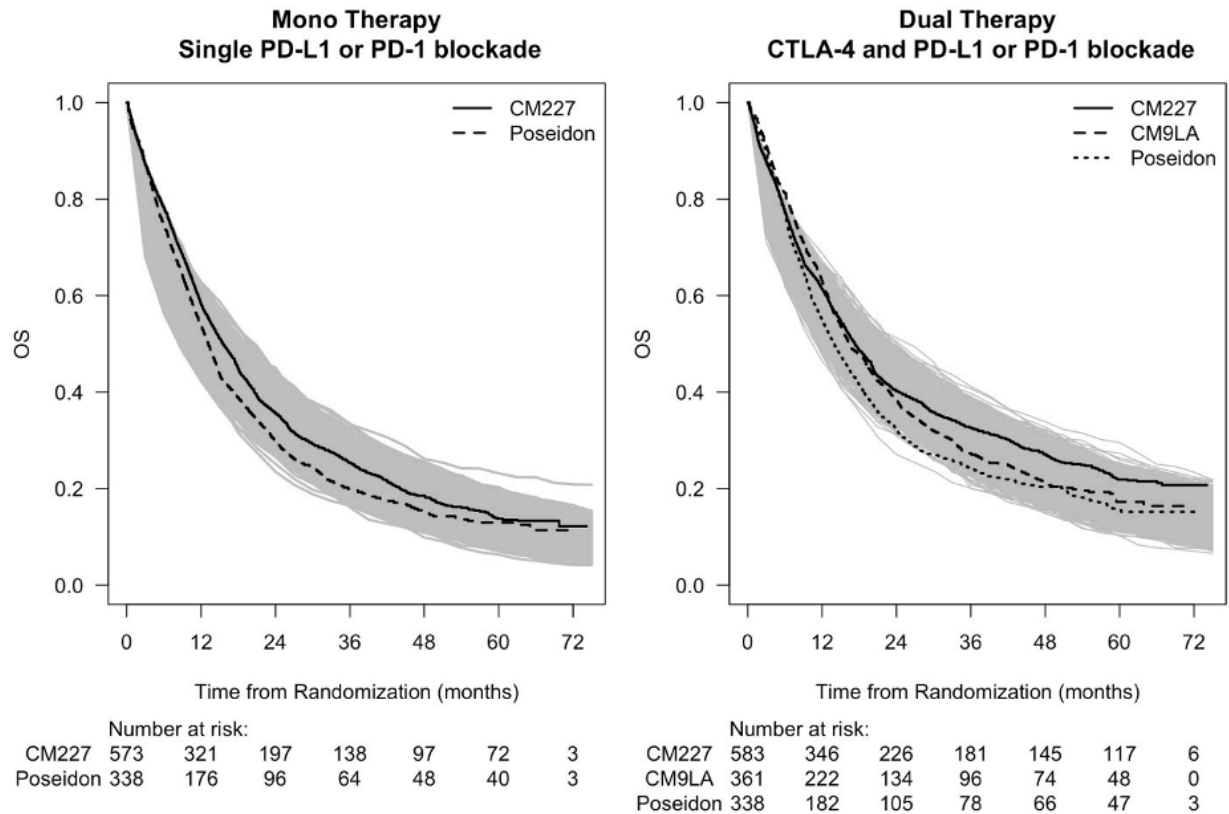

Figure 3: The projected survival curves associated with the mono- and dural inhibitor arms of a new trial, assuming it has a baseline covariates profile similar to that of trials in the mixed histology cluster (POSEIDON, CM227 and CM9LA). Each grey curve is a possible survival curve that might be observed in the future trial, given published Kaplan-Meier curves (colored black) for the mono therapies (left panel) and the dual therapies (right panel) from trials in the mixed histology cluster.

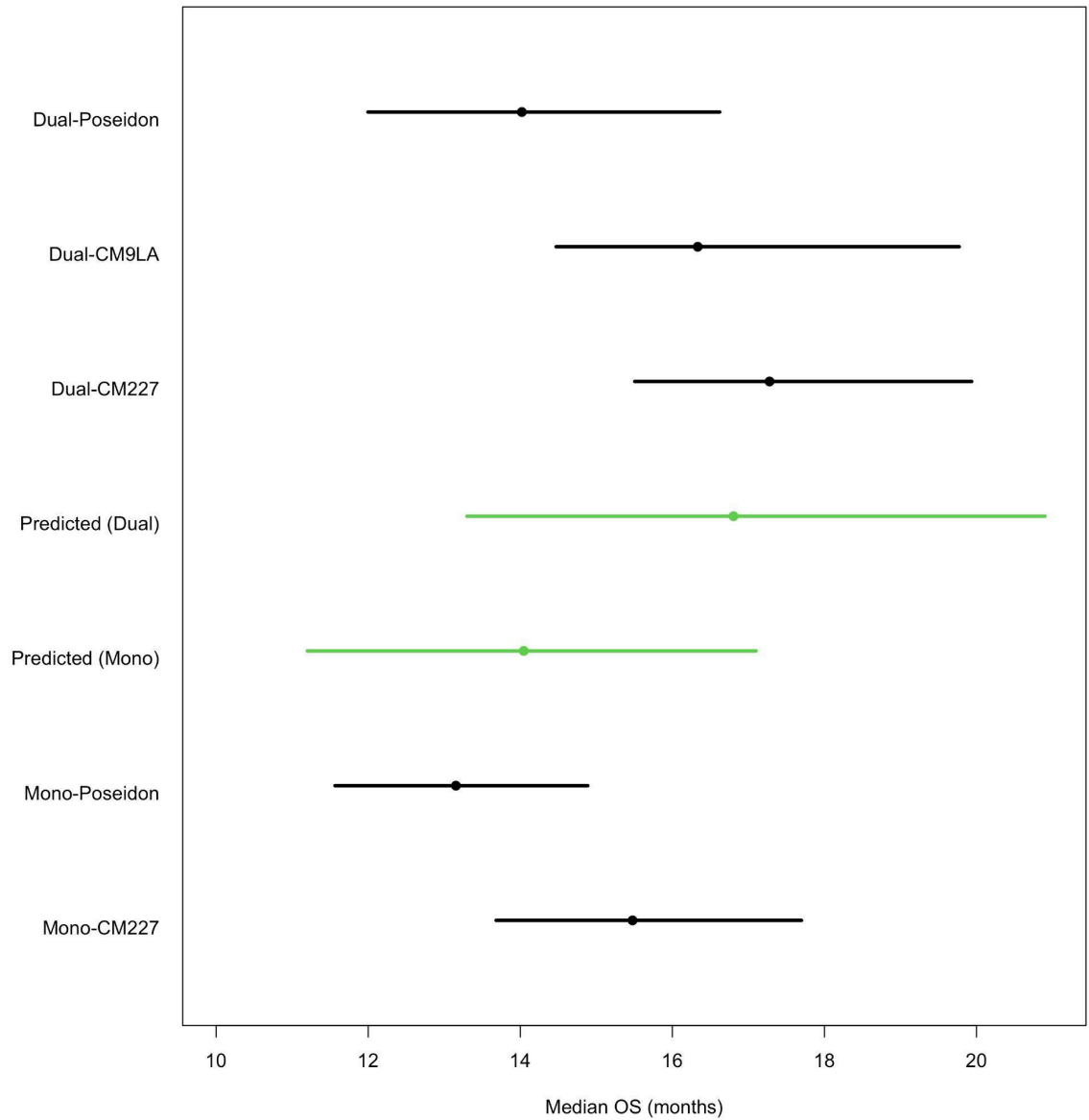

Figure 4: The model-predicted median OS and 95% CIs for mono- and dual- ICIs in a new trial with patient profile similar to POSEIDON, CM227 and CM9LA. The Kaplan-Meier estimator for the median of each trial is shown in black.

# Supplementary Appendix: Technical Details

February 9, 2026

## Contents

# 1 IPD Reconstruction

We are interested in synthesizing evidence on survival outcomes of patients treated with the same treatment or the same class of treatments $\mathcal{C}$. Suppose after a thorough literature review and a rigorous selection process, we have identified $J$ different arms consisting of patients treated by $\mathcal{C}$ from historical trials sharing key eligibility criteria. We denote these historical arms as $H_1, \ldots, H_J$, with the corresponding sample size $n_1, \ldots, n_J$.

Kaplan-Meier (KM) curves of previous studies are commonly published along with a risk table, which displays the number of patients at risk at an equally spaced time grid. Taking the risk table and the extracted XY coordinates of KM curves as input, the modified-iKM method[1] generates individual patient data (IPD) for each $H_j$, where $j = 1, \ldots, J$. Let $i$ index patient. Following the standard survival analysis notation, we denote $Y_i$ as the follow up time for this patient and let $\delta_i = 1$ if the event of interest is observed and $\delta_i = 0$ if it is censored. We denote the reconstructed IPD for $H_j$ as $D_j = \{(Y_i, \delta_i); i = 1, \ldots, n_j\}$.

# 2 Survival Evidence Synthesis

We model the reconstructed IPD using the beta–Stacy process (BSP), which is conjugate to right-censored observations [2]. We begin with a brief review of the BSP; for a detailed treatment, see Walker and Muliere [2].

## 2.1 A Review of beta-Stacy Process

Let $F(t)$ denote the unknown cumulative distribution function (CDF) of discretized event times defined on a prespecified time grid $\{t_1, \ldots, t_K\}$. Define the increment of CDF at $t_k$ as

$$\theta_k = F(t_k) - F(t_{k-1}), \ t_0 = 0, \ F(0) = 0.$$

Conditional on $\theta_1, \ldots, \theta_{k-1}$, the random increment $\theta_k$ follows a beta-Stacy distribution [3] with density function given by

$$f(\theta_k | \alpha_k, \beta_k, \theta_1, \ldots, \theta_{k-1}) = \frac{\Gamma(\alpha_k + \beta_k)}{\Gamma(\alpha_k)\Gamma(\beta_k)} \theta_k^{\alpha_k - 1} \frac{(1 - \theta_1 - \cdots - \theta_k)^{\beta_k - 1}}{(1 - \theta_1 - \cdots - \theta_{k-1})^{\alpha_k + \beta_k - 1}},$$

where $\alpha_k > 0$ and $\beta_k > 0$.

As a specific case, $\theta_1$ has a beta distribution with parameters $\alpha_1$ and $\beta_1$.

We can construct a prior distribution for $F$ by defining it as the sum of random increments

$$F(t_k) = \sum_{j=1}^{k} \theta_j, \qquad (2.1)$$

assuming each $\theta_j$ follows a beta-Stacy distribution with parameters $\alpha_j$ and $\beta_j$ conditional on pre-vious increments of the CDF.

Following Walker and Muliere[2], the resulting prior distribution for $F(t)$ is a discrete-time beta–Stacy process (BSP) defined on $\{t_1, \ldots, t_K\}$, which is denoted as

$$F(t) \sim BSP(\{\alpha_k, \beta_k\}).$$

The BSP prior is conjugate to right-censored survival data. Let $d_k$ and $r_k$ represent the number of events and the number of patients at risk at $t_k$, respectively. Denote $m_k = r_k - d_k$. The posterior distribution of $F(t)$ is again a BSP with updated parameters $\{\alpha_k^*, \beta_k^*\}$ and jumps at $\{t_k\}$. We express this posterior distribution as

$$F(t)|\boldsymbol{m}, \boldsymbol{d} \sim BSP(\{\alpha_k^*, \beta_k^*\}), \ \alpha_k^* = \alpha_k + d_k, \ \beta_k^* = \beta_k + m_k, \tag{2.2}$$

where $\boldsymbol{m} = (m_1, \ldots, m_K)$ and $\boldsymbol{d} = (d_1, \ldots, d_K)$.

Let $G(t)$ represent a prespecified survival function and let $c(t)$ denote a known positive function representing the strength of prior knowledge at time $t$. Following [2], we parameterize the BSP prior as

$$\alpha_k = c(t_k)\{G(t_{k-1}) - G(t_k)\}, \tag{2.3}$$

$$\beta_k = c(t_k)G(t_k). \tag{2.4}$$

After reparameterization, the prior distribution of $S(t) = 1 - F(t)$ is centered at $G(t)$. That is,

$$E\{S(t)|G(t), c(t)\} = G(t).$$

## 2.2 A Hierarchical Bayesian Survival Model

For each historical arm $H_j$, let $F_j$ denote the CDF of discretized survival times defined on a prespecified time grid $\boldsymbol{\tau}^* = (t_1, \ldots, t_K)$. That is, we assume that the discretized IPD can only take values from $\boldsymbol{\tau}^*$. The discretization of reconstructed survival times $Y_1, \ldots, Y_{n_j}$ is necessary to account for the reduced accuracy of the reconstructed data compared to the original data. Risk tables from different trials may have time grid of varying temporal resolution. To minimize information loss due to discretization, we recommend choosing $\boldsymbol{\tau}^*$ as the time grid with the best temporal resolution among the $J$ historical arms. Although $t_K$ can technically go up to infinity, we consider $t_K$ as the maximum follow up times of different trials. For example, if one historical trial has a risk table with a time grid $(6, 12, \ldots, 60)$ and the other historical trial has a risk table with a time grid $(3, 6, \ldots, 72)$, we will choose the latter as $\boldsymbol{\tau}^*$.

Let $F^*$ denote the CDF of patients treated by $\mathcal{C}$ in a new trial. We build a Bayesian hierarchical model (BHM), assuming CDFs from different trials follow a discrete-time BSP prior:

$$F^*, F_1, \ldots, F_J \sim BSP(\{\alpha_k, \beta_k\}), \tag{2.5}$$

defined on the grid $\boldsymbol{\tau}^*$.

Denote $\eta_{jk}$ the hazard rate for $H_j$ at $t_k$. We can relate the hazard rates to $F_j$ through the product-limit estimator

$$F_j(t) = 1 - S_j(t) = 1 - \prod_{t_k \leq t}(1 - \eta_{jk}). \tag{2.6}$$

Based on Walker and Muliere[2], Equation 2.5 implies the hazard rates from different trials have independently distributed Beta priors

$$\eta_k^*, \eta_{1k}, \ldots, \eta_{Jk} \sim \text{Beta}(\alpha_k, \beta_k). \tag{2.7}$$

Let $d_{jk}$ and $r_{jk}$ represent the number of events and the number of patients at risk at $t_k$ for $k = 1, \ldots, K$ based on the discretized IPD from $H_j$. At each $t_k$, we consider a binomial model

$$d_{jk} \sim \text{Bin}(r_{jk}, \eta_{jk}). \tag{2.8}$$

Following Walker and Muliere[2], the parameters $\{\alpha_k, \beta_k\}$ from different time points can be linked by a survival model $G(t)$ and a precision function $c(t)$. We define

$$\alpha_k = c(t_k)\{G(t_{k-1}) - G(t_k)\},$$
$$\beta_k = c(t_k)G(t_k).$$

Recent empirical work using 500 trial arms (approximately 220,000 events) shows that survival in oncology trials is well described by two-parameter Weibull models[4]. Thus, we consider

$$G(t) = exp(-\lambda t^\kappa).$$

Unlike previous works assuming $G(t)$ and $c(t)$ as fixed[2, 5, 6], we consider they are both random and specify independent vague priors for $\lambda$ and $\kappa$, where

$$\lambda \sim Gamma(0.01, 0.01), \tag{2.9}$$

and

$$\kappa \sim Gamma(0.01, 0.01). \tag{2.10}$$

Setting $c(t)$ equal to a positive constant $c$, it follows $\beta_k = \beta_{k-1} - \alpha_k$ and the number of patients at risk at $t = 0$ is $\alpha_1 + \beta_1 = cG(0) = c$. Therefore, we can interpret $c$ as the prior effective sample size (ESS) of the BSP prior.

The total sample size of the historical patients is $N = \sum_{j=1}^{J} n_j$. As $c$ is the ESS of the BSP prior, $c$ should not exceed $N$ and a reasonable prior for $c$ is

$$\frac{c}{N} \sim \text{Beta}(1, 1). \tag{2.11}$$

## 3 Predictive Distribution for Survival Curves in a New Trial

Using the hierarchical BSP model described in Section 2.2, the prior predictive distribution of the survival function in a new trial can be induced based on

$$S^*(t) = \prod_{t_k \leq t} (1 - \eta_k^*), \tag{3.1}$$

which accounts for the uncertainties in estimating both $G(t)$ and $c$.

We approximate the predictive distribution of $S^*(t)$ using another beta–Stacy process and assume

$$S^*(t) \sim BSP(\{\alpha_k^*, \beta_k^*\}),$$

with parameters matched to the usual BSP parametrization in (2.3):

$$\alpha_k^* = c^* \{G^*(t_{k-1}) - G^*(t_k)\},$$

and

$$\beta_k^* = c^* G^*(t_k),$$

where

$$G^*(t) = \exp\{-\lambda^* t^{\kappa^*}\}, \tag{3.2}$$

and $\lambda^*$ and $\kappa^*$ are the posterior means of $\lambda$ and $\kappa$.

Walker and Muliere [2] showed the hazard rates associated with survival functions following a beta-Stacy process are independently distributed beta random variables. Thus, we can assume the hazard rates in a new trial $\eta_1^*, \ldots, \eta_K^*$ are independently distributed as $\text{Beta}(\alpha_1^*, \beta_1^*), \ldots, \text{Beta}(\alpha_K^*, \beta_K^*)$, then the prior variance of $S^*(t)$ can be calculated in closed form:

$$V^*(t) = \prod_{\tau_k \leq t} u_k^* - \prod_{\tau_k \leq t} v_k^*, \tag{3.3}$$

where

$$v_k^* = \left(\frac{\alpha_k^*}{\alpha_k^* + \beta_k^*}\right)^2,$$

and

$$u_k^* = v_k^* + \frac{\alpha_k^* \beta_k^*}{(\alpha_k^* + \beta_k^*)^2 (\alpha_k^* + \beta_k^* + 1)},$$

Let $\hat{V}(t)$ denote the estimated variance of $S^*(t)$ based on a large sample from the predictive distribution of $S^*(t)$ given by (3.1). Given $\lambda^*$ and $\kappa^*$, we solve $c^*$ by finding the $c^*$ that minimizes

$$\sum_{k=1}^{K} (V^*(t_k) - \hat{V}(t_k))^2. \tag{3.4}$$

We can define the effective sample size (ESS) and effective number of events (ENE) of the prior predictive distribution as

$$ENE = \sum_{k=1}^{K} \alpha_k^*, \ ESS = c^*. \tag{3.5}$$